\documentclass[a4paper,11pt]{article}
\usepackage{pos}
\usepackage{slashed}
\usepackage{float}
\usepackage{sectsty}

\newcommand{\beq}{\begin{equation}}
\newcommand{\eeq}{\end{equation}}
\newcommand{\bea}{\begin{eqnarray}}
\newcommand{\eea}{\end{eqnarray}}
\newcommand{\nn}{\nonumber}

\title{G-structures for AdS$_2$ solutions}
 \ShortTitle{G-structures for AdS$_2$ solutions}

\author*[a, b]{Niall T. Macpherson}
\author[c]{Achilleas Passias}

\affiliation[a]{Department of Physics, University of Oviedo,\\
  Avda. Federico Garcia Lorca s/n, 33007 Oviedo, Spain}

\affiliation[b]{Instituto Universitario de Ciencias y Tecnolog\'ias Espaciales de Asturias (ICTEA),\\
Calle de la Independencia 13, 33004 Oviedo, Spain}

\affiliation[c]{Department of Physics, University of Ioannina,\\ GR45110, Ioannina, Greece}

\emailAdd{macphersonniall@uniovi.es}
\emailAdd{achilleas.passias@phys.ens.fr}

\abstract{We report on the classification of supersymmetric AdS$_2$ solutions of Type II supergravity and the discovery of new solutions.}

\FullConference{Proceedings of the Corfu Summer Institute 2024 "School and Workshops on Elementary Particle Physics and Gravity" (CORFU2024)\
12 - 26 May, and 25 August - 27 September, 2024\\
Corfu, Greece\\}

\begin{document}
\maketitle

\section{Introduction}
Quantum gravity in spacetimes that asymptote two-dimensional anti-de Sitter spacetime (AdS$_2$) is a highly interesting problem, owning to AdS$_2$ arising as the near-horizon geometry of extremal black holes and the AdS$_2$/CFT$_1$ correspondence not being as well-understood as its higher-dimensional counterparts, due to exotic features such as disconnected boundaries of the spacetime. It thus worth exploring the space and properties of AdS$_2$ solutions of string theory.

Our focus is on supersymmetric AdS$_2$ solutions: supersymmetry provides a 
technically simplifying assumption in solving the equations of motion and at the same time it provides computational control on both sides of the AdS$_2$/CFT$_1$ correspondence. For example, the entropy of asymptotically anti-de Sitter black holes was obtained from the Witten index of the dual quantum mechanics, computed using supersymmetric localization techniques \cite{Benini:2015eyy}. The high dimensionality of the internal space of AdS$_2$ solutions of string theory allows for large space of solutions. We approach this problem by classifying the possible geometries according to the G-structures and the constraints they satisfy, as dictated by supersymmetry. 

In particular, we have imposed $\mathcal{N} = 1$ supersymmetry: such solutions generically support an SU(3)-structure on their internal manifold M$_8$, which can experience an enhancement to a G$_2$-structure. In \cite{Legramandi:2023fjr}, we performed an SU(3)-structure torsion classes analysis and expressed the fluxes and other physical fields in terms of these, in general (we omit these results from the present report). Based on these results, we have derived two new classes of AdS$_2$ solutions. In (massive) Type IIA supergravity we derive an $\mathcal{N} = 1$ supersymmetric class for which M$_8$ is a warped product of a weak G$_2$-manifold and an interval and which is locally defined in terms of a degree three polynomial. In Type IIB supergravity we find a class of AdS$_2 \times$ S$^2 \times$ CY$_2 \times \Sigma_2$ solutions governed by a harmonic function on $\Sigma_2$ and partial differential equations reminiscent of D3-D7-brane configurations.

\section{G-structures for minimally supersymmetric AdS$_2$ in Type II supergravity}\label{sec:sec2}
By definition, an AdS$_2$ solution of Type II supergravity can be decomposed as
\begin{align}
ds^2&=e^{2A}ds^2(\text{AdS}_2)+ds^2(\text{M}_8),\nn\\[2mm]
F_{\pm}&= f_{\pm}+e^{2A}\text{vol}(\text{AdS}_2)\wedge\star_8 \lambda(f_{\pm}),~~~~ H=e^{2A}\text{vol}(\text{AdS}_2)\wedge H_1+H_3,
\end{align}
where $(e^{2A},H_1,H_3,f_{\pm})$ and the dilaton $\Phi$ have support on the internal manifold M$_8$ and we take AdS$_2$ to have inverse radius $m$. $F_{\pm}$ is the democratic RR polyform such that 
\beq
F_+= F_0+ F_2+F_4+F_6+F_8+F_{10},~~~~~F_-=F_1+F_3+F_5+F_7+F_9;
\eeq
here and elsewhere the upper/lower signs should be taken in Type IIA/IIB. Finally, acting on a $k$-form we have that $\lambda(C_k)=(-1)^{\lfloor\frac{k}{2}\rfloor}C_k$ which appears in $F_{\pm}$ such that it obeys the self duality constraint 
\beq
\star_{10}\lambda(F_{\pm})=F_{\pm},
\eeq
giving it the correct number of degrees of freedom.

As shown in \cite{Legramandi:2023fjr} (which builds on the foundations of \cite{Tomasiello:2011eb}), Type II AdS$_2$ solutions that preserve ${\cal N}=1$ supersymmetry, \textit{i.e.} two real supercharges, generically do so  in terms of an SU(3)-structure\footnote{Strictly speaking this is for AdS$_2$ solutions that do not experience a necessary enhancement to at least AdS$_3$ locally.} in $d=8$. This comes equipped with a real 2-form $J$, a holomorphic 3-form $\Omega$ and two vielbein directions $(U,V)$ such that
\begin{align}
&J\wedge \Omega=0,~~~J\wedge J\wedge J=\frac{3}{4}i\Omega\wedge\overline{\Omega},~~~\iota_U(J,~\Omega)=\iota_V(J,~\Omega)=0\nn\\[2mm]
&ds^2(\text{M}_8)= ds^2(\text{M}_{\text{SU}(3)})+V^2+U^2,~~~~\text{vol}(\text{M}_8)=\frac{1}{3!}J\wedge J\wedge J\wedge U\wedge V,
\end{align}
with $(J,\Omega)$ spanned by the vielbein directions on $\text{M}_{\text{SU}(3)}$. Together with two point dependent phases $(\alpha,\beta)$ these can be used to define two polyforms on M$_8$, namely $(\psi_{\pm},\psi_{\mp})$ which decompose into their parts with legs parallel and orthogonal to $V$ as 
\begin{align}
\psi_{\pm}&=\frac{ 1}{16}\text{Re}\bigg[\psi^{(7)}_{\pm}+\cos\beta\psi^{(7)}_{\mp}\wedge V \bigg],~~~~
\psi_{\mp}=\pm \frac{1}{16}\sin\beta\text{Re}\bigg[ \text{Re}\psi^{(7)}_{\mp}\bigg],
\end{align}
where $\sin\beta=0$ is incompatible with AdS$_2$. The $d=7$ polyform themselves decompose into parts orthogonal and parallel to $U$ in terms of SU(3)-structure polyforms as 
\begin{align}
\psi^{(7)}_{\pm}&=\psi^{\text{SU(3)}}_{\pm}+i \psi^{\text{SU(3)}}_{\mp}\wedge U,\\[2mm]
\psi^{\text{SU(3)}}_+&=e^{i\alpha} e^{-i J},~~~~\psi^{\text{SU(3)}}_-=\Omega.\nn
\end{align}
Though this clearly gives rise to an SU(3)-structure on M$_8$ generically, there is actually an enhancement to a G$_2$-structure when $e^{i\alpha}=1$, where $(J,\Omega,U)$ arrange themselves into the associated real 3-form as
\beq
\Phi_3=-(J\wedge U+\text{Re}\Omega),~~~~\iota_{V}\star_8\Phi_3=\frac{1}{2} J\wedge J- U\wedge \text{Im}\Omega.
\eeq

Necessary and sufficient conditions for supersymmetry can be phrased in purely geometric terms as differential conditions relating the internal supergravity data $(e^{2A},\Phi,H_1,H_3,f_{\pm})$ to the G-structure forms in terms of $(\beta,V,\psi_{\pm},\psi_{\mp})$. These conditions take the concise from
\begin{subequations}\label{eq:BPS}
\begin{align}
&e^{2A}H_1=me^{A} \sin\beta V-d(e^{2A}\cos\beta),~~~~d(e^{A}\sin\beta V)=0,\label{eq:BPS1}\\[2mm]
&d_{H_3}(e^{A-\Phi}\psi_{\pm})=\pm \frac{1}{16}e^{A}\sin\beta V\wedge f_{\pm}\label{eq:BPS2},\\[2mm]
&d_{H_3}(e^{2A-\Phi}\psi_{\mp})- m e^{A-\Phi}\psi_{\pm}=\mp\frac{1 }{16}e^{2A}(\star_8\lambda f_{\pm}+\cos\beta f_{\pm}),\label{eq:BPS3}\\[2mm]
&(\psi_{\pm},f_{\pm})_8=\pm\frac{1}{4} e^{-\Phi}\left(me^{-A}-\frac{1}{2}\sin\beta \iota_{V} H_1\right)\text{vol}(\text{M}_8),\label{eq:BPS4}
\end{align}
\end{subequations}
where the bracket $(X,Y)_8$ is defined as the 8-form part of $X\wedge \lambda(Y)$.

While the above conditions ensure supersymmetry, they do not ensure a solution of Type II supergravity. For that, one must of course solve the equations of motion and Bianchi identities of the fluxes. However, it is generically true that supersymmetry and some subset of the field equations imply what remains, and AdS$_2$ is no exception. Leveraging integrability arguments from \cite{Legramandi:2018qkr} it is possible to establish that all the Type II field equations are implied if \eqref{eq:BPS} hold and
\beq
dH_3=0,~~~~ \iota_V(d_{H_3} f_{\pm})= 0,~~~~\cos\beta \bigg[d(e^{-2\Phi}\star_8 H_1)+\frac{1}{2}(f_{\pm},f_{\pm})_8\bigg]=0,\label{eq:solconds}
\eeq
are imposed away from localised sources, where possible delta function corrections can appear.

\section{${\cal N}=1$ supersymmetric class containing a weak G$_2$-manifold}
In \cite{Dibitetto:2018gbk}, a class of ${\cal N}=8$ supersymmetric AdS$_2$ solutions in massive Type IIA supergravity was constructed and further studied in \cite{Lozano:2025ief}. This class is a warped product of $\text{AdS}_2\times \text{S}^7$ and an interval, defined in terms of a (local) degree three polynomial of the interval coordinate. It can be realised either as a near-horizon limit of the D0-F1-D8 brane intersection of \cite{Imamura:2001cr}, or as a double analytic continuation of the AdS$_7$ solutions of \cite{Apruzzi:2013yva}. Through the observation that the 7-sphere supports a weak G$_2$-holonomy, it is possible to use \eqref{eq:BPS} to construct a generalisation of this class preserving generically ${\cal N}=1$ supersymmetry. Here we summarise the class, details on its construction can be found in \cite{Legramandi:2023fjr}.

The class is a warped product of $\text{AdS}_2\times\text{M}_{\text{WG}_2}\times I$,  for  $\text{M}_{\text{WG}_2}$ any weak G$_2$-manifold, that is defined in terms of two functions $(h,v)$ of $r$ which spans the interval. Its NSNS sector takes the following form 
\begin{align}
\frac{ds^2}{L^2}&=\sqrt{\frac{h}{h''}}\bigg[\frac{h h''\sqrt{1-7v}}{8\Delta}ds^2(\text{AdS}_2)+ \bigg(\frac{h''}{8h\sqrt{1-7v}}d{\rho}^2+\frac{\sqrt{1-7v}}{(v-1)^2}ds^2(\text{M}_{\text{WG}_2})\bigg)\bigg],\nn\\[2mm]
H&=\frac{L^2}{8\sqrt{2}}d\left(\frac{hh'(1-7v)}{\Delta}-\rho\right)\wedge\text{vol}(\text{AdS}_2),~~~~e^{-\Phi}= \frac{\sqrt{\Delta}(1-v)^{\frac{7}{2}}}{c_0 L^3 (1-7v)^{\frac{5}{4}}}\left(\frac{h''}{h}\right)^{\frac{3}{4}},\nn\\[2mm]
\Delta&=2h h''-(1-7v)(h')^2,
\end{align}
where $(L,c_0)$ are constants.

To express the RR sector it is important to appreciate that a defining property of a weak G$_2$-manifold is the existence of a G$_2$-structure 3-form on $\text{M}_{\text{WG}_2}$, $\Phi_{\text{WG}_2}$, obeying the relation
\beq
d\Phi_{\text{WG}_2}= 4 \star_{\text{WG}_2}\Phi_{\text{WG}_2}.
\eeq
Generically, this can appear in the RR fluxes, though its non-closure implies that it cannot appear in the NSNS 3-form. In terms of $(h,v,\Phi_{\text{WG}_2})$ the magnetic part of the RR fluxes take the following form.
\begin{align}
f_0&=\frac{4}{c_0L^4}\partial_{\rho}\left(\frac{(1-v)^{\frac{7}{2}}h''}{(1-7 v)}\right),\nn\\[2mm]
f_4&=-\frac{2v^{\frac{3}{4}}}{c_0(1-v)(1-7v)}\partial_{\rho}\left((1-v)^{\frac{3}{2}}v^{\frac{1}{4}}h'\right) \Phi_{\text{WG}_2}\wedge d\rho,\nn\\[2mm]
f_6&=\frac{4\sqrt{v}}{c_0}\partial_{\rho}\left(\frac{h \sqrt{v}}{\sqrt{1-v}}\right) \star_{\text{WG}_2}\Phi_{\text{WG}_2}\nn,\\[2mm]
f_8&=\frac{L^4}{c_0}\left(\partial_{\rho}\left(\frac{(1-7 v)h h'}{(1-v)^{\frac{7}{2}}h''}\right)-\frac{h (3-7 v)}{(1-v)^{\frac{7}{2}}}\right) \text{vol}(\text{M}_{\text{WG}_2})\wedge d\rho.
\end{align}
The above definitions ensure at least ${\cal N}=1$ supersymmetry, but to have a solution one must also impose the following partial differential equations, away from possible sources, 
\begin{align}
&\partial^2_{\rho}\left(\frac{(1-v)^{\frac{7}{2}}h''}{(1-7 v)}\right)=0,\nn\\
&\partial_{\rho}\left(\sqrt{v}\partial_{\rho}\left(\frac{h \sqrt{v}}{\sqrt{1-v}}\right)\right)-\frac{2v^{\frac{3}{4}}}{(1-v)(1-7v)}\partial_r\left((1-v)^{\frac{3}{2}}v^{\frac{1}{4}}h'\right)=0,\label{eq:definingPDES}
\end{align}
which imply \eqref{eq:solconds}.

This class permits $\text{M}_{\text{WG}_2}$ to be any weak G$_2$-manifold, but for AdS/CFT considerations one would like this to be compact. Compact examples include, but are not limited to,  foliations over nearly-K\"{a}hler bases $\text{M}_{\text{NK}}$ of the form
\beq
ds^2(\text{M}_{\text{WG}_2})=d\theta^2+\sin^2\theta ds^2(\text{M}_{\text{NK}}),
\eeq
where the known compact examples with closed form metrics are $\text{M}_{\text{NK}}=(\text{S}^6,\,\text{S}^3\times \text{S}^3,\,\mathbb{CP}^3,\,\mathbb{F}^3)$. Generically, this leads to a solution bounded between conical G$_2$ singularities, the exception being for S$^6$ which is when $\text{M}_{\text{WG}_2}=\text{S}^7$. For such manifolds the G$_2$-structure forms decompose as
\begin{align}
\Phi_{\text{WG}_2}&=\sin^2\theta d\theta\wedge J_{\text{NK}}+\sin^3\theta \text{Re}(e^{-i\theta}\Omega_{\text{NK}}),\nn\\
\star_{\text{WG}_2}\Phi_{\text{WG}_2}&= -\frac{1}{2}\sin^4 \theta J^2_{\text{NK}}+ \sin^3\theta d\theta\wedge \text{Im}(e^{-i\theta}\Omega_{\text{NK}}),
\end{align}
where $(J_{\text{NK}},\Omega_{\text{NK}})$ define the nearly-K\"{a}hler structure, and so obey
\beq
dJ_{\text{NK}}=3 \text{Re}\Omega_{\text{NK}},~~~~d\text{Im}\Omega_{\text{NK}}=-2J_{\text{NK}}^2.
\eeq

The conditions \eqref{eq:definingPDES} appear too complicated to expect a general closed form solution, however they truncate considerably when one assumes that $v=v_0$ is a constant. Then one finds that they reduce to 
\beq
h''''=0,~~~~ v_0(1+5 v_0)=0,\label{eq: simplesol}
\eeq
which implies that $h$ is locally a degree three polynomial and leads to two branches with
\beq
v_0=0,~~~~v_0=-\frac{1}{5}.
\eeq
For $v_0=0$ one has $f_4=f_6=0$, so $\Phi_{\text{WG}_2}$ drops out of the fluxes entirely and if we select the 7-sphere as the weak G$_2$-manifold we reproduce the class of \cite{Dibitetto:2018gbk} and ${\cal N}=1$ is enhanced to ${\cal N}=8$. When $v_0=-\frac{1}{5}$,  $(f_2,f_4)$ remain non-trivial so even when $\text{M}_{\text{WG}_2}=\text{S}^7$, the presence of $\Phi_{\text{WG}_2}$ stops an enhancement of supersymmetry from ${\cal N}=1$.

While the class of solution with $v_0=0$ was to be expected, the class with $v_0=-\frac{1}{5}$ is quite novel and opens up some interesting possibilities. First, we find that, as with the case $v_0=0$, solutions are defined by the ODE
\beq
h'''\sim F_0,
\eeq
but $F_0$ need only be constant piecewise with its discontinuities indicating source D8-branes along the interior of the interval spanned by $r$. For the class with $v_0=0$, considered in its 7-sphere avatar in \cite{Lozano:2025ief}, this was shown to allow for a broad class of solutions for which the interval was bounded between physical behaviours, either regular or singular. We expect that the same can be achieved with $v_0=-\frac{1}{5}$. Second, if one takes $\text{M}_{\text{WG}_2}=\text{S}^7$ one can perform a double analytic continuation to arrive at AdS$_7$ in massive Type IIA. By performing the same procedure with $v_0=-\frac{1}{5}$ it should be possible to derive a deformed AdS$_7$ solution that could have interesting implications for AdS$_7$/CFT$_6$.

Finally, let us stress that \eqref{eq: simplesol} is a simple way to solve \eqref{eq:definingPDES}; we do not expect it to be the general solution. Indeed, it may be possible to uncover more interesting solutions, at least numerically.

\section{${\cal N}=4$ supersymmetric $\text{AdS}_2\times\text{S}^2\times \text{CY}_2\times\Sigma_2$}\label{sec:section4}
Famously, the near-horizon limit of D1-D5 branes wrapping a Calabi-Yau 2-fold gives rise to $\text{AdS}_3\times \text{S}^3\times \text{CY}_2$, where a compact solution requires that either CY$_2=\mathbb{T}^4$ or $K_3$. This solution preserves small ${\cal N}=(4,4)$ superconformal symmetry, but half of this is preserved if one T-dualises on both the Hopf fiber U(1)'s of AdS$_3$ and S$^3$. The result of doing this is an $\text{AdS}_2\times\text{S}^2\times \text{S}^1\times \text{S}^1\times  \text{CY}_2$ solution in Type IIB, which provides an uplift of the near-horizon limit of the extremal Reissner--Nordstrom black hole into string theory. On the other hand, there have been several works \cite{Lozano:2019emq,Lozano:2019jza,Lozano:2019zvg,Lozano:2019ywa,Couzens:2021veb,Lima:2022hji} that generalise the D1-D5 near-horizon and its Type IIA analogue to include warp factors for additional branes that are backreacted on the CY$_2$ and break supersymmetry to small ${\cal N}=(4,0)$. These allow one to construct compact solutions bounded between D-brane and O-plane singularities even when the CY$_2$ is not itself compact. This begs the question, what is the most general class of solutions containing a warped $\text{AdS}_2\times\text{S}^2\times\text{CY}_2$ factor in Type IIB? Such a class should be of interest both to the study of black holes and the holographic description of superconformal quantum mechanics. In this section we present this class; for the sake of brevity we again direct the interested reader to \cite{Legramandi:2023fjr} for details on its construction.

We find a class that is a foliation of $\text{AdS}_2\times\text{S}^2\times \text{CY}_2$ over a Riemann surface, it  preserves small ${\cal N}=4$ superconformal symmetry and its NSNS sector is given by
\begin{align}
ds^2&=  \frac{u}{\sqrt{h_3 h_7}}\left(\frac{1}{\Delta_2}ds^2(\text{AdS}_2)+\frac{1}{\Delta_1}ds^2(\text{S}^2)\right)+\sqrt{\frac{h_3}{h_7}}ds^2(\text{CY}_2)+\frac{\sqrt{h_3h_7}}{u}(dx_1^2+dx_2^2)\bigg],\\
e^{-\Phi}&=c_0 \sqrt{\Delta_1\Delta_2} h_7,~~~~ H= dB_0\wedge \text{vol}(\text{AdS}_2)+d\tilde{B}_0\wedge\text{vol}(\text{S}^2)+ dx_1\wedge X^{(1,1)}_1+ dx_2\wedge X^{(1,1)}_2,\nn
\end{align}
where CY$_2$ can be any Calabi-Yau 2-fold and to ease the presentation we have introduced the following functions
\beq
\Delta_1=1+ \frac{(\partial_{x_1}u)^2}{h_3 h_7},~~~~\Delta_2=1- \frac{(\partial_{x_2}u)^2}{h_3 h_7},~~~~B_0=-x_2-\frac{u \partial_{x_2}u}{ h_3 h_7\Delta_2},~~~~\tilde{B}_0=x_1-\frac{u \partial_{x_1}u}{ h_3 h_7\Delta_1}.
\eeq
In the above $c_0$ is a constant, while $(h_7,u)$ are functions that depend only on $(x_1,x_2)$, while $h_3$ has support on $(x_1,x_2)$ and the coordinates on CY$_2$. On the other hand, the 2-forms $(X^{(1,1)}_1,X^{(1,1)}_2)$ are real primitive $(1,1)$-forms with legs on CY$_2$ but which can have functional dependence on $(x_1,x_2,\text{CY}_2)$; what this means is that they are constrained to obey
\beq
J\wedge X^{(1,1)}_a=\Omega\wedge X^{(1,1)}_a=0,~~~~a=1,2,
\eeq
where $(J,\Omega)$ are the closed SU(2)-structure forms defining the specific CY$_2$ which in our conventions obey
\beq
J\wedge\Omega=0,~~~~ J\wedge J=\frac{1}{2}\Omega\wedge \overline{\Omega}, ~~~~\star_4 J=J,~~~~\star_4\Omega =\Omega,
\eeq
for $\star_4$ the Hodge star operator on the unwarped CY$_2$ manifold. This means that each of $X^{(1,1)}_a$ can be expanded in the basis of the three independent anti-self-dual 2-forms on CY$_2$, so that they are each fixed up to three real functions of $(x_1,x_2,\text{CY}_2)$. The magnetic portion of the RR fluxes is given by 
\begin{align}
f_1&=c_0\left(\star_2 d_2h_7+d\left(\frac{\partial_{x_1}u\partial_{x_2}u}{h_3}\right)\right),\nn\\[2mm]
f_3&= \tilde{B}_0f_1\wedge \text{vol}(\text{S}^2)-c_0\bigg[\bigg(x_1\star_2 d_2h_7+ h_7dx_2-d\left(\frac{\partial_{x_2}u(u-x_1\partial_{x_1}u)}{h_3}\right)\bigg)\wedge\text{vol}(\text{S}^2)\nn\\[2mm]
&+h_7\left(dx_1\wedge X^{(1,1)}_2-dx_2\wedge X^{(1,1)}_1\right)+\frac{\partial_{x_1}u\partial_{x_2}u}{h_3}\left( dx_1\wedge X^{(1,1)}_1+ dx_2\wedge X^{(1,1)}_2\right)\bigg],\nn\\[2mm]
f_5&= \tilde{B}_0f_3\wedge \text{vol}(\text{S}^2)-c_0\bigg[\frac{h_7}{u}\star_4 d_4 h_3\wedge \text{vol}_2+ \star_2 d_2 h_3\wedge\text{vol}_4+ d\left(\frac{\partial_{x_1}u\partial_{x_2}u}{h_7}\right)\wedge \text{vol}_4\nn\\[2mm]
&\bigg(-x_1 h_7\left(dx_1\wedge X_2^{(1,1)}-dx_2\wedge X_1^{(1,1)}\right)+\frac{\partial_{x_2}u(u-x_1 \partial_{x_1}u)}{h_3}\left(dx_1\wedge X_1^{(1,1)}+dx_2\wedge X_2^{(1,1)}\right)\bigg)\wedge \text{vol}(\text{S}^2)\bigg],\nn\\[2mm]
f_7&= \tilde{B}_0f_5\wedge \text{vol}(\text{S}^2)+c_0\bigg[-d\left(\frac{\partial_{x_2}u(u-x_1\partial_{x_1}u)}{h_7}\right)\wedge \text{vol}_4\nn\\[2mm]
&+x_1\frac{h_7}{u}\star_4 d_4 h_3\wedge \text{vol}_2+(x_1 \star_2d_2 h_3+h_3 dx_2)\wedge \text{vol}_4\bigg]\wedge \text{vol}(\text{S}^2),
\end{align}
where we have decomposed
\beq
d=d_2+ d_{4},~~~~ d_2= dx_i\wedge  \partial_{x_i},
\eeq
for $d_{4}$ the exterior derivative on CY$_2$. We also take  $(\text{vol}_2,\text{vol}_4)$ to be the volume forms on the unwarped $(x_1,x_2)$ and CY$_2$ directions and $\star_2$ to be the Hodge star operator on the unwarped Riemann surface.

Unlike our previous example, in this case supersymmetry does not only fix the form of the physical fields in terms of some functions and their derivatives, it also demands that we impose one additional constraint, namely that 
\beq
\square_2u=0,~~~~\square_2 = \partial_{x_1}^2+\partial_{x_2}^2, \label{eq:pdesusy}
\eeq
which much hold globally, \textit{i.e} it is not possible to add delta function sources to this expression. When this is imposed, supersymmetry is ensured, but an on-shell solution is not; that requires that away from sources \eqref{eq:solconds} is imposed. This, first constrains the primitive $(1,1)$-forms such that
\beq
d_4 X^{(1,1)}_1=d_4 X^{(1,1)}_2=0,~~~~\partial_{x_2}X^{(1,1)}_1=\partial_{x_1}X^{(1,1)}_2,
\eeq
and second leads to the following system of partial differential equations
\begin{align}
 \square_2 h_7=0,~~~~  \frac{h_7}{u}\square_4 h_3+\square_2  h_3+ h_7\left((X^{(1,1)}_1)^2+(X^{(1,1)}_2)^2\right)=0,\label{eq:bianchis}
\end{align}
where $\square_4$ is the Laplacian on CY$_2$. These generalise (in terms of $(u,X^{(1,1)}_a)$) the equations defining localised D3-branes inside the worldvolume of localised D7-branes in flat space \cite{Youm:1999ti} --- this is the reason for the numerical subscripts on $(h_3,h_7)$. We note also that if we fix $du=0$ then $(h_3,h_7)$ appear precisely where one would expect the warp factors of D3-branes wrapped on $\text{AdS}_2\times \text{S}^2$ and D7-branes wrapped on $\text{AdS}_2\times \text{S}^2\times \text{CY}_2$ to appear. We should stress that \eqref{eq:bianchis} should hold away from the loci of sources, contrary to \eqref{eq:pdesusy}: these can have delta function sources appearing on the right-hand side as long as they take a form consistent with the appropriate extended object, \textit{i.e.} D3/D7-branes or O3/O7-planes.

The class we have derived is quite broad and contains several others in certain limits. If one imposes that $\partial_{x_1}$ is an isometry of the entire background and $X^{(1,1)}_1=0$ one recovers the class considered in \cite{Lozano:2021rmk}, which is itself a double analytic continuation of a class in \cite{Lozano:2020txg}. Conversely, when one fixes $X^{(1,1)}_2=0$ and makes $\partial_{x_2}$ an isometry one generates the result of T-dualising the $\text{AdS}_3\times \text{S}^2\times \text{CY}_2\times I$ class of \cite{Lozano:2019emq} on the Hopf fiber of AdS$_3$. Both \cite{Lozano:2021rmk} and \cite{Lozano:2020txg} contain solutions that lie within our broader class. They assume however that the warp factors do not depend on the CY$_2$ directions and that $X_{1}^{(1,1)}=X_{2}^{(1,1)}=0$. As such they are both contained in an $\text{AdS}_2\times \text{S}^2\times \text{CY}_2\times \Sigma_2$ class derived and studied in \cite{Chiodaroli:2009yw,Chiodaroli:2009xh}. This should also be contained within our class in the limit in which $X_{1}^{(1,1)}=X_{2}^{(1,1)}=0$ and $h_3=h_3(x_1,x_2)$, where the system is now defined in terms of three harmonic functions.

The general class of solutions we have derived have relative warping between the AdS$_2$ and S$^2$ directions that depends on the CY$_2\times \Sigma_2$ directions, while the near-horizon of extremal Reissner--Nordstrom is round $\text{AdS}_2\times \text{S}^2$. As such, only the $u= \text{constant}$ limit of our class, where $\Delta_1=\Delta_2=1$, provides an embedding of this solution into ten dimensions. Nonetheless, even under this restriction, the class is quite broad and so promises to provide interesting uplifts of this black hole near-horizon into string theory. It is also likely that the $\text{AdS}_2\times \text{S}^2$ factor in this limit can be replaced with any solution of ${\cal N}=2$ minimal supergravity in four dimensions. This would provide an embedding for the entire extremal black hole which is a supersymmetric solution of this 4d theory \cite{Romans:1991nq}.

\section{Outlook}
We have classified minimally supersymmetric AdS$_2$ solutions of Type II supergravity in terms of SU(3)- and G$_2$-structures on the internal manifold M$_8$. Based on this classification, we have constructed new families of solutions: in (massive) Type IIA supergravity a family for which M$_8$ contains a weak G$_2$-manifold and which is locally defined in terms of a degree three polynomial and in Type IIB supergravity a family of AdS$_2 \times$ S$^2 \times$ CY$_2 \times \Sigma_2$ solutions governed by a harmonic function on $\Sigma_2$ and partial differential equations reminiscent of D3-D7-brane configurations.

These constructions have spurred further studies: in \cite{Lozano:2025ief}, members of the Type IIA family of solutions for which the weak G$_2$-manifold is an S$^7$ or its orbifold S$^7/\mathbb{Z}_k$ were analysed and proposals for the dual field theories were put forward based on brane configurations. These members have enhanced $\mathcal{N}=8$, for the S$^7$, and $\mathcal{N}=6$, for its orbifold, supersymmetry. It would be interesting to make contact with the recent developments on the construction of the superconformal mechanics \cite{Krivonos:2024zcj,Krivonos:2025dzw}. 

Beyond this, the geometric conditions for supersymmetry presented in section \ref{sec:sec2} have already been exploited to construct a broad class of  $\text{AdS}_2\times \mathbb{CP}^3\times \Sigma_2$ solutions in Type IIB  \cite{Conti:2025djz} whose existence was postulated in \cite{Conti:2023rul}. This class of solutions likewise preserve ${\cal N}=6$, but instead of a degree three polynomial, their local form is defined in terms of two holomorphic functions. In this sense, the class is an AdS$_2$ analogue of \cite{DHoker:2007zhm} and likewise promises to contain numerous solutions of physical interest.

Together with \cite{Hong:2019wyi}, which derived G-structures conditions for AdS$_2$ solutions in $d=11$, our results fully geometrize the problem of constructing AdS$_2$ solutions of string theory. There are many interesting avenues to pursue with these new tools. For instance:

\begin{itemize}
\item ${\cal N}=8$ (sixteen real supercharges) is maximal for AdS$_2$ solutions \cite{Gran:2017qus}. Such solutions promise to realise the avatars of the duality between AdS$_2$ and superconformal quantum mechanics that are under the most control. As such, it would be very interesting to classify the maximally supersymmetric AdS$_2$ solutions. 
\item There should be AdS$_2$ classes that contain the holographic description of line defects in higher-dimensional CFTs, waiting to be uncovered. A particularly interesting avenue to pursue in this regard would be line defects in ABJM which is an active area of study, see for instance \cite{Drukker:2019bev,Gorini:2020new,Castiglioni:2023uus}.
\item Another very interesting avenue to explore would be a full classification of charged extremal black hole near-horizons in ten and eleven dimensions. We give part of this in section \ref{sec:section4}, but this should be far from the full picture.  Constructing the full picture would serve as a stepping stone to embed the full black hole geometries into string theory.
\end{itemize}
Finally, while the results of \cite{Legramandi:2023fjr} and \cite{Hong:2019wyi} do provide useful tools to embed charged black hole near-horizons into string theory, we should stress that real black holes also rotate. To capture the near-horizons of such black holes in their extremal limit requires one to generalise to solutions which contain an AdS$_2$ factor with a U(1) fibered over it. There has been some progress in this direction \cite{Couzens:2020jgx}, but it would be interesting to pursue a more systematic study of such squashed AdS$_3$ solutions.


\begin{thebibliography}{99}
\bibitem{Benini:2015eyy}
F.~Benini, K.~Hristov and A.~Zaffaroni,
``Black hole microstates in AdS$_{4}$ from supersymmetric localization,''
JHEP \textbf{05} (2016), 054
[arXiv:1511.04085 [hep-th]].

\bibitem{Legramandi:2023fjr}
A.~Legramandi, N.~T.~Macpherson and A.~Passias,
``G-structures for black hole near-horizon geometries,''
JHEP \textbf{06} (2024), 056
[arXiv:2309.01714 [hep-th]].

\bibitem{Tomasiello:2011eb}
A.~Tomasiello,
``Generalized structures of ten-dimensional supersymmetric solutions,''
JHEP \textbf{03} (2012), 073
[arXiv:1109.2603 [hep-th]].

\bibitem{Legramandi:2018qkr}
A.~Legramandi, L.~Martucci and A.~Tomasiello,
``Timelike structures of ten-dimensional supersymmetry,''
JHEP \textbf{04} (2019), 109
[arXiv:1810.08625 [hep-th]].

\bibitem{Dibitetto:2018gbk}
G.~Dibitetto and A.~Passias,
``AdS$_{2}$ \texttimes{} S$^{7}$ solutions from D0-F1-D8 intersections,''
JHEP \textbf{10} (2018), 190
[arXiv:1807.00555 [hep-th]].

\bibitem{Lozano:2025ief}
Y.~Lozano, N.~T.~Macpherson and A.~Passias,
``On $\text{AdS}_2\times \text{S}^7$, its $\mathbb{Z}_k$ orbifold and their dual quantum mechanics,''
[arXiv:2503.23227 [hep-th]].

\bibitem{Imamura:2001cr}
Y.~Imamura,
``1/4 BPS solutions in massive IIA supergravity,''
Prog. Theor. Phys. \textbf{106} (2001), 653-670
[arXiv:hep-th/0105263 [hep-th]].

\bibitem{Apruzzi:2013yva}
F.~Apruzzi, M.~Fazzi, D.~Rosa and A.~Tomasiello,
``All AdS$_7$ solutions of type II supergravity,''
JHEP \textbf{04} (2014), 064
[arXiv:1309.2949 [hep-th]].

\bibitem{Lozano:2019emq}
Y.~Lozano, N.~T.~Macpherson, C.~Nunez and A.~Ramirez,
``AdS$_3$ solutions in Massive IIA with small $\mathcal{N}=(4,0)$ supersymmetry,''
JHEP \textbf{01} (2020), 129
[arXiv:1908.09851 [hep-th]].

\bibitem{Lozano:2019jza}
Y.~Lozano, N.~T.~Macpherson, C.~Nunez and A.~Ramirez,
``1/4 BPS solutions and the AdS$_3$/CFT$_2$ correspondence,''
Phys. Rev. D \textbf{101} (2020) no.2, 026014
[arXiv:1909.09636 [hep-th]].

\bibitem{Lozano:2019zvg}
Y.~Lozano, N.~T.~Macpherson, C.~Nunez and A.~Ramirez,
``Two dimensional ${\cal N}=(0,4)$ quivers dual to AdS$_3$ solutions in massive IIA,''
JHEP \textbf{01} (2020), 140
[arXiv:1909.10510 [hep-th]].

\bibitem{Lozano:2019ywa}
Y.~Lozano, N.~T.~Macpherson, C.~Nunez and A.~Ramirez,
``AdS$_3$ solutions in massive IIA, defect CFTs and T-duality,''
JHEP \textbf{12} (2019), 013
[arXiv:1909.11669 [hep-th]].

\bibitem{Couzens:2021veb}
C.~Couzens, Y.~Lozano, N.~Petri and S.~Vandoren,
``N=(0,4) black string chains,''
Phys. Rev. D \textbf{105} (2022) no.8, 086015
[arXiv:2109.10413 [hep-th]].

\bibitem{Lima:2022hji}
M.~Lima, N.~T.~Macpherson, D.~Melnikov and L.~Ypanaque,
``On generalised D1-D5 near horizons and their spectra,''
JHEP \textbf{04} (2023), 060
[arXiv:2211.02702 [hep-th]].

\bibitem{Youm:1999ti}
D.~Youm,
``Partially localized intersecting BPS branes,''
Nucl. Phys. B \textbf{556} (1999), 222-246
[arXiv:hep-th/9902208 [hep-th]].

\bibitem{Lozano:2021rmk}
Y.~Lozano, C.~Nunez and A.~Ramirez,
``$\text{AdS}_2\times \text{S}^2\times \text{CY}_2$ solutions in Type IIB with 8 supersymmetries,''
JHEP \textbf{04} (2021), 110
[arXiv:2101.04682 [hep-th]].

\bibitem{Lozano:2020txg}
Y.~Lozano, C.~Nunez, A.~Ramirez and S.~Speziali,
``New AdS$_{2}$ backgrounds and $ \mathcal{N} $ = 4 conformal quantum mechanics,''
JHEP \textbf{03} (2021), 277
[arXiv:2011.00005 [hep-th]].

\bibitem{Chiodaroli:2009yw}
M.~Chiodaroli, M.~Gutperle and D.~Krym,
``Half-BPS Solutions locally asymptotic to AdS(3) x S**3 and interface conformal field theories,''
JHEP \textbf{02} (2010), 066
[arXiv:0910.0466 [hep-th]].

\bibitem{Chiodaroli:2009xh}
M.~Chiodaroli, E.~D'Hoker and M.~Gutperle,
``Open Worldsheets for Holographic Interfaces,''
JHEP \textbf{03} (2010), 060
[arXiv:0912.4679 [hep-th]].

\bibitem{Romans:1991nq}
L.~J.~Romans,
``Supersymmetric, cold and lukewarm black holes in cosmological Einstein-Maxwell theory,''
Nucl. Phys. B \textbf{383} (1992), 395-415
[arXiv:hep-th/9203018 [hep-th]].

\bibitem{Krivonos:2024zcj}
S.~Krivonos and A.~Nersessian,
``N=8 superconformal mechanics: Direct construction,''
Phys. Lett. B \textbf{863} (2025), 139373
[arXiv:2411.18345 [hep-th]].

\bibitem{Krivonos:2025dzw}
S.~Krivonos and A.~Nersessian,
``Two faces of $N=7,8$ superconformal mechanics,''
[arXiv:2504.13651 [hep-th]].

\bibitem{Conti:2025djz}
A.~Conti, Y.~Lozano and N.~T.~Macpherson,
``$\mathcal{N}=6$ supersymmetric AdS$_2 \times \mathbb{CP}^3\times \Sigma_2 $,''
[arXiv:2503.23585 [hep-th]].

\bibitem{Conti:2023rul}
A.~Conti,
``AdS3 T duality and evidence for N=5,6 superconformal quantum mechanics,''
Phys. Rev. D \textbf{108} (2023) no.12, 126007
[arXiv:2306.09139 [hep-th]].

\bibitem{DHoker:2007zhm}
E.~D'Hoker, J.~Estes and M.~Gutperle,
``Exact half-BPS Type IIB interface solutions. I. Local solution and supersymmetric Janus,''
JHEP \textbf{06} (2007), 021
[arXiv:0705.0022 [hep-th]].

\bibitem{Hong:2019wyi}
J.~Hong, N.~T.~Macpherson and L.~A.~Pando Zayas,
``Aspects of AdS$_{2}$ classification in M-theory: solutions with mesonic and baryonic charges,''
JHEP \textbf{11} (2019), 127
[arXiv:1908.08518 [hep-th]].

\bibitem{Gran:2017qus}
U.~Gran, J.~Gutowski and G.~Papadopoulos,
``All superalgebras for warped AdS$_2$ and black hole near horizon geometries,''
Class. Quant. Grav. \textbf{36} (2019) no.23, 235009
[arXiv:1712.07889 [hep-th]].

\bibitem{Drukker:2019bev}
N.~Drukker, D.~Trancanelli, L.~Bianchi, M.~S.~Bianchi, D.~H.~Correa, V.~Forini, L.~Griguolo, M.~Leoni, F.~Levkovich-Maslyuk and G.~Nagaoka, \textit{et al.}
``Roadmap on Wilson loops in 3d Chern\textendash{}Simons-matter theories,''
J. Phys. A \textbf{53} (2020) no.17, 173001
[arXiv:1910.00588 [hep-th]].

\bibitem{Gorini:2020new}
N.~Gorini, L.~Griguolo, L.~Guerrini, S.~Penati, D.~Seminara and P.~Soresina,
``The topological line of ABJ(M) theory,''
JHEP \textbf{06} (2021), 091
[arXiv:2012.11613 [hep-th]].

\bibitem{Castiglioni:2023uus}
L.~Castiglioni, S.~Penati, M.~Tenser and D.~Trancanelli,
``Wilson loops and defect RG flows in ABJM,''
JHEP \textbf{06} (2023), 157
[arXiv:2305.01647 [hep-th]].

\bibitem{Couzens:2020jgx}
C.~Couzens, E.~Marcus, K.~Stemerdink and D.~van de Heisteeg,
``The near-horizon geometry of supersymmetric rotating AdS$_{4}$ black holes in M-theory,''
JHEP \textbf{05} (2021), 194
[arXiv:2011.07071 [hep-th]].

\end{thebibliography}
\end{document}